\documentstyle[11pt]{article}
\begin{document}
\date{} 
\begin{flushright}
{MIT-CTP-3178}\\
{YITP-01-57}
\end{flushright}
\vspace{2cm}
\begin{center}
{\Large {\bf On Relation between Two Models of\\
Gauge Field Localization on a Brane}
\vskip0.8truein
{\large Motoi Tachibana}~\footnote{E-mail:{\tt motoi@lns.mit.edu}
~~JSPS Research Fellow\\
on leave from Yukawa Institute for Theoretical Physics,
Kyoto University, Kyoto 606-8502, Japan}\\}
\vskip0.4truein
\centerline{ {\it 
Center for Theoretical Physics,
Massachusetts Institute of Technology, Cambridge, MA 02139, USA.}}
\end{center}
\vskip0.8truein
\centerline{\bf Abstract}
\vskip0.3truein
We discuss the relation between two different models which are
recently proposed as the model of localizing bulk gauge fields 
on a brane. In the former model, the localization of gauge field
is achieved by adding both bulk and boundary mass terms while
in the latter, it is done by taking into consideration the
coupling between the gauge field and the dilaton field (this
model is also regarded as the gauge theory with nontrivial
dielectric \lq \lq constant''). We make a certain
transformation for the gauge field in the latter Lagrangian.
As the result, we find those two models are closely related
to each other.
\vskip0.13truein
\baselineskip=0.5 truein plus 2pt minus 1pt
\baselineskip=18pt
\addtolength{\parindent}{2pt}
\newpage
Since Randall and Sundrum proposed a new solution to the
hierarchy problem\cite{rs1,rs2}, theories with (infinite)
extra spatial dimensions have received much attention. Many 
people have tried to extend the Randall-Sundrum model in different
fashions. One of what they have attempted is to put the
standard model into the bulk, i.e., they considered bulk
fields(scalar, fermion and gauge fields) as well as 
gravitons. Bulk scalar field was originally introduced
by Goldberger and Wise to stabilize the size of the extra 
dimension\cite{GW}. Bulk fermion fields in the Randall-Sundrum model
have been considered as well\cite{fermi} and it has been shown that they 
could be localized on both positive and negative tension
branes by intoducing the kink-like mass term\cite{mass}. 

On the other hand, bulk gauge field in the Randall-Sundrum
model has a very different story\footnote{The first field theory 
example of gauge field localization on a brane was 
suggested in Ref.\cite{dv}}. Unlike the scalar and
fermion fields, it has been wellknown that the zero mode of 
the bulk gauge field is not localized on a brane\cite{gauge}. 
Actually it is flat in the extra dimension. The reason consists in
the rescaling property of the gauge field action as has
already been addressed by several authors\cite{conformal}. 
The 4-dimensional kinetic term of the gauge field is not warped.

However there have more recently appeared several models 
of gauge field localization in the Randall-Sundrum geometry
with infinite extra space dimension. In this paper,
we shall restrict our consideration into the following two
different models. One is introduced by the authors of \cite{ghoroku,kogan} 
where the localization is achived by adding both bulk and boundary
mass terms of the gauge field. The other one is introduced by  
\cite{kehagias} where the localization is achieved by taking into
account the coupling between the gauge field and the dilaton
field. Both models have the gauge field zero mode localized
on a brane whose wave function has the peak just on the
brane. 

In the light of these situations, it is interesting to investigate
the relation between these two models of gauge field localization.
In fact, it is the purpose of this paper. 

Let us start with introducing Lagrangians of the two models.
The action with both bulk and boundary mass terms is given
by\cite{ghoroku,kogan}
\begin{equation}
S_{massive} = \int d^5 X\sqrt{-G}\biggl(-\frac{1}{4}G^{MN}G^{PQ}
F_{MP}F_{NQ}-\frac{1}{2}(M^2+c\delta(y))G^{MN}A_M A_N\biggr),
\label{lag1}
\end{equation}
where $M, N, P, Q = 0, 1, 2, 3, y$ and $F_{MN}=\partial_M A_N
-\partial_N A_M$. $A_M(x^{\mu},y)$ is the bulk U(1) gauge field.
$G_{MN}$ is the 5-dimensional metric defined by
\begin{equation}
ds^2 = e^{-2|y|/L}\eta_{\mu \nu}dx^{\mu}dx^{\nu}+dy^2.
\label{metric}
\end{equation}
Here the parameters $M$ and $c$ denote the bulk and the
boundary masses, respectively. They are supposed to 
appear through spontaneous breaking of the gauge invariance
in the bulk\cite{kogan}. 

On the other hand, the action with the coupling between
the gauge field and the dilaton field is given by\cite{kehagias}
\begin{equation}
S_{dilaton} = \int d^5 X\sqrt{-G}\biggl(-\frac{1}{4}
e^{-\lambda \pi/2\sqrt{3M^3_{Pl}}}G^{MN}G^{PQ}F_{MP}F_{NQ}
+\frac{1}{2}(\partial \phi)^2+\frac{1}{2}(\partial \pi)^2
-V(\phi, \pi)\biggr),
\label{lag2}
\end{equation}
where $\lambda$ is a dimensionless coupling depending on the
underlying theory. The scalar field $\phi$ is the stuff the
domain wall is made off while the another scalar field $\pi$
is nothing but the dilaton field. $V(\phi, \pi)$ is supposed to be
usual Higgs-type potential. In the absence of the U(1) gauge 
field, under the assumption of the following form of the metric, 
\begin{equation}
ds^2 = e^{-2A(y)}\eta_{\mu \nu}dx^{\mu}dx^{\nu}+dy^2,
\label{metric2}
\end{equation}
we find the solutions of the equations of motion(we assume
both $\phi$ and $\pi$ depend on only the fifth dimensional
coordinate $y$): 
\begin{eqnarray}
\phi(y) &=& v\tanh(a y), \nonumber \\
A(y) &=& -\beta \ln\cosh^2(a y) -\frac{\beta}{2}\tanh^2(a y), \nonumber\\
\pi(y) &=& \beta\sqrt{3M^3_{Pl}}\biggl(
\ln\cosh^2(a y) +\frac{1}{2}\tanh^2(a y)\biggr),
\label{solution}
\end{eqnarray}
where $a^{-1}$ is the \lq \lq width'' of the domain wall which 
is given in terms of some parameters appearing in the original action.
$\beta$ is given by $v^2/36M^3_{Pl}$ where $v$ is the expectation
value of the scalar field. $M_{Pl}$ is the 5-dimensional Planck scale.

Obviously we have to take the brane limit, i.e., $a \rightarrow
\infty$ to compare these two models appropriately. This limit
is taken with the quantity $\xi \equiv 3\beta a$ fixed. As the 
result, we obtain
\begin{eqnarray}
\phi(y) &=& v\theta(y)+\cdots, \nonumber \\
A(y) &=& -\frac{2{\xi}^2}{3}|y|-\frac{{\xi}^2}{6a}(1-4\ln2)+\cdots,
\nonumber \\
\pi(y) &=& \sqrt{3M^3_{Pl}}\biggl(
\frac{2{\xi}^2}{3}|y|+\frac{{\xi}^2}{6a}(1-4\ln2)+\cdots\biggr),
\label{solution2}
\end{eqnarray}
where $\theta(y)$ is the step function defined by
\begin{eqnarray}
\theta(y) =\left\{\begin{array}{ll}
+1 & {\rm for} \quad y > 0 \\
-1 & {\rm for} \quad y < 0
\end{array}\right.
\label{step}
\end{eqnarray}
We are ready to go further.  As has been suggested, 
we expect that there is some relation between two models described 
by the Lagrangians (\ref{lag1}) and (\ref{lag2}).  Here we shall 
give the following conjecture;
\begin{equation}
e^{-\lambda \pi/4\sqrt{3M^3_{Pl}}}A^{(dilaton)}_M = A^{(massive)}_M,
\label{conjecture}
\end{equation}
where $A^{(massive)}_M$ and $A^{(dilaton)}_M$ denote the gauge fields
appearing in the Lagrangians (\ref{lag1}) and (\ref{lag2}), respectively.

Now we shall examine the above conjecture by direct computation.
Let us first define the function $\epsilon(\pi)$ as
\begin{equation}
e^{-\lambda \pi/4\sqrt{3M^3_{Pl}}} \equiv {\epsilon}^2(\pi).
\label{dielectric}
\end{equation}
It is interesting to note here that the function
$\epsilon(\pi)$ (correctly speaking, ${\epsilon}^2$) can be regarded
as the dielectric \lq \lq constant''. In Ref.\cite{kawai}, a similar
model has been discussed to show confinement in the bulk via the
dielectric effect proposed in \cite{thooft} with an appropriate
choice for $\epsilon$ and $V$.

It is straightforward to see that
\begin{eqnarray}
\epsilon(\pi)F_{MP} &=& \epsilon(\partial_M A_P-\partial_P A_M), \nonumber \\
&=& \partial_M(\epsilon A_P)-\partial_P(\epsilon A_M)
      -(\partial_M \epsilon \cdot A_P-\partial_P \epsilon \cdot A_M).
\label{henkei1}
\end{eqnarray}
Here we put $\epsilon(\pi)A_P(x,y) = \tilde{A}_P(x,y)$.
It is clear that
\begin{equation}
\epsilon(\pi)F_{MP}(A) = F_{MP}(\tilde{A})-
\biggl(\frac{\partial_M \epsilon}{\epsilon}\tilde{A}_P
-\frac{\partial_P \epsilon}{\epsilon}\tilde{A}_M \biggr).
\label{henkei2}
\end{equation}
Therefore we have
\begin{eqnarray}
{\epsilon}^2(\pi)G^{MN}G^{PQ}F_{MP}(A)F_{NQ}(A)
&=& G^{MN}G^{PQ}F_{MP}(\tilde{A})F_{NQ}(\tilde{A}) \nonumber \\
& &  -2G^{MN}G^{PQ}F_{MP}(\tilde{A})\biggl(
\frac{\partial_N \epsilon}{\epsilon}\tilde{A}_Q
-\frac{\partial_Q \epsilon}{\epsilon}\tilde{A}_N\ \biggr) \nonumber \\
& & +2G^{MN}G^{PQ}
\biggl(\frac{\partial_M \epsilon}{\epsilon}\frac{\partial_N \epsilon}{\epsilon}
\tilde{A}_P\tilde{A}_Q-
\frac{\partial_M \epsilon}{\epsilon}\frac{\partial_Q \epsilon}{\epsilon}
\tilde{A}_P\tilde{A}_N \biggr).
\label{henkei3}
\end{eqnarray}
From below, we work with the gauge condition
\begin{equation}
A_y(x^{\mu}, y) = 0,
\label{gauge}
\end{equation}
where $\mu =  0,1,2,3$.

Then it is simple to show the following equation
\begin{eqnarray}
{\epsilon}^2(\pi)G^{MN}G^{PQ}F_{MP}(A)F_{NQ}(A)
&=& G^{MN}G^{PQ}F_{MP}(\tilde{A})F_{NQ}(\tilde{A}) \nonumber \\
& & -4G^{\mu \nu}\biggl(\frac{\partial_y \epsilon}{\epsilon}\biggr)
\tilde{A}_{\mu}\partial_y \tilde{A}_{\nu}
+2G^{\mu \nu}\biggl(\frac{\partial_y \epsilon}{\epsilon}\biggr)^2
\tilde{A}_{\mu}\tilde{A}_{\nu}.
\label{henkei4}
\end{eqnarray}
Plugging (\ref{henkei4}) into the Lagrangian
(\ref{lag2}), we are led to the following expression
(the scalar parts are dropped);
\begin{eqnarray}
S_{dilaton} &=& \int d^5 X \sqrt{-G}\Biggl[
-\frac{1}{4}G^{MN}G^{PQ}F_{MP}(\tilde{A})F_{NQ}(\tilde{A}) \nonumber \\
& & \qquad \qquad \qquad \qquad
+G^{\mu \nu}\biggl(\frac{\partial_y \epsilon}{\epsilon}\biggr)
\tilde{A}_{\mu}\partial_y \tilde{A}_{\nu}
-\frac{1}{2}G^{\mu \nu}\biggl(\frac{\partial_y \epsilon}{\epsilon}\biggr)^2
\tilde{A}_{\mu}\tilde{A}_{\nu} \Biggr ].
\label{action1}
\end{eqnarray}
By combining the equation (\ref{dielectric}) with the asymptotic form of
$\pi(y)$ at $a \rightarrow \infty$,  we can evaluate the term
$\partial_y \epsilon/\epsilon \equiv d \ln\epsilon(\pi)/dy$ as follows;
\begin{eqnarray}
\frac{d \ln \epsilon(\pi)}{dy} &=& 
-\frac{\lambda}{4\sqrt{3M^3_{Pl}}}\frac{d\pi}{dy}, \nonumber \\
&=&-\frac{\lambda {\xi}^2}{6}\theta(y) + {\cal O}(1/a).
\label{log}
\end{eqnarray}
Again, $\theta(y)$ is the step function defined through eq.(\ref{step}).

Thus we obtain
\begin{eqnarray}
S_{dilaton}&=& \int d^5 X \sqrt{-G}\Biggl[
-\frac{1}{4}G^{MN}G^{PQ}F_{MP}(\tilde{A})F_{NQ}(\tilde{A}) \nonumber \\
& & \qquad \qquad \qquad \qquad
-\frac{1}{2}{M'}^2 G^{\mu \nu}\tilde{A}_{\mu}\tilde{A}_{\nu}
-\frac{\lambda {\xi}^2}{6}G^{\mu \nu}\theta(y)
\tilde{A}_{\mu}\partial_y \tilde{A}_{\nu} \Biggr ],
\label{action2}
\end{eqnarray}
where 
\begin{equation}
{M'}^2 = \frac{{\lambda}^2{\xi}^4}{36},
\label{M}
\end{equation}
which plays a role of the bulk mass of the gauge field. 

Here we make partial integration for the third term in 
the action (\ref{action2}). The result is that  
\begin{eqnarray}
S_{dilaton}[\tilde{A}]&=& \int d^5 X \sqrt{-G}\Biggl[
-\frac{1}{4}G^{MN}G^{PQ}F_{MP}(\tilde{A})F_{NQ}(\tilde{A}) \nonumber \\
& & -\frac{1}{2}({M'}^2+c'\delta(y))G^{\mu \nu}\tilde{A}_{\mu}\tilde{A}_{\nu}
+\frac{\lambda {\xi}^2}{6}\frac{1}{\sqrt{-G}}\theta(y)
\partial_y(\sqrt{-G}G^{\mu \nu}\tilde{A}_{\mu})\tilde{A}_{\nu} \Biggr ],
\label{action3}
\end{eqnarray}
where the coefficient in front of the delta function of the second term 
is defined by 
\begin{equation}
c' = -\frac{2\lambda {\xi}^2}{3},
\label{c}
\end{equation}
which corresponds to the mass parameter of the gauge field on the brane 
at $y=0$. This is also interpreted as the interaction one 
between the bulk gauge field and the brane. Note here that the sign of
$c'$ is negative as long as $\lambda$ is positive. So we have tachyonic
mass term on the brane in this case.

Eq.(\ref{action3})  has the same form as eq.(\ref{lag1}) except
the third term  of the right hand side when
we choose $A_y =0$ gauge in eq.(\ref{lag1}).  
As has been wellknown, the field transformation such as
eq.(\ref{conjecture}) is used to generate the mass term 
via the kinetic term. That was one of our original motivations
against the conjecture (\ref{conjecture}). As we see from
the process of deriving eq.(\ref{action3}), the origin of
the boundary mass term, which is proportional to the delta
function, consists in the orbifold geometry of 
the Randall-Sundrum model.   

Finally let us consider the parameters appearing in both 
Lagrangians in detail.  
As we see from eq.(\ref{M}) and eq.(\ref{c}) immediately,
\begin{equation}
M' \approx |c'|.
\label{mc1}
\end{equation}
On the other hand, The parameters appearing in the action (\ref{lag1}), $M$
and $c$, are related through the boundary condition on the brane at $y=0$
as follows;
\begin{equation}
|c| = \frac{2}{L}(\sqrt{1+M^2 L^2}-1) \approx M^2 L,
\label{mc2}
\end{equation}
where $L$ denotes the radius of 5-dimensional AdS space. In the Randall-Sundrum
model, $L^{-1}$ is of the order of the 5-dimensional Planck scale, i.e., 
$L^{-1} \approx M_{Pl}$.  Therefore we find $|c| \approx 
\frac{M}{M_{Pl}}M$ in this case. From the viewpoint of solving the hierarchy
problem, the bulk mass parameter $M$ should be of the order of $M_{Pl}$.
If it is applied, the boundary mass parameter $|c|$ is of the order of
$M_{Pl}$ as well and is given as $|c| \approx M$.
As the result, we obtain the same relation between
the bulk and the boundary mass parameters shown in eq.(\ref{mc1}).

To conclude, in this paper, we discussed the relation between 
recently proposed two models of gauge field localization on a brane 
in the Randall-Sundrum model.  We made a certain transformation of 
the gauge field and it was shown that those two models are closely 
(not exactly) related by direct computation. One important property
to derive the mass term on the brane was the orbifold geometry
of the Randall-Sundrum model. We examined the parameters
appearing in Lagrangians of the two models. As the result, we obtained
the equivalent relations between the bulk and the boundary mass parameters
in both models, which was done through a natural requirement 
to solve the hierarchy problem. The physical meaning of the conjecture 
(\ref{conjecture}) is, however, still less clear even if it
generates both bulk and boundary mass terms. This will be a future 
issue.

\vspace{0cm}

\begin{center}
{\large{\bf Acknowledgments}}
\end{center}
This work was supported in part by a Grant-in-Aid for Scientific 
Research from Ministry of Education, Science, Sports and Culture 
of Japan (No.~3666).



\begin{thebibliography}{99}
\bibitem{rs1} 
L. Randall and R. Sundrum, 
{\it \lq \lq A Large Mass Hierarchy from a Small Extra Dimension,''}
Phys. Rev. Lett. {\bf 83}~(1999)~3370, {\tt hep-ph/9905221}.
\bibitem{rs2} 
L. Randall and R. Sundrum, 
{\it \lq \lq An Alternative to Compactification,''}
Phys. Rev. Lett. {\bf 83}~(1999)~4690, {\tt hep-th/9906064}.
\bibitem{GW} 
W. Goldberger and M. Wise, 
{\it \lq \lq Bulk Fields in the Randall-Sundrum Scenario,''}
Phys. Rev. {\bf D60}~(1999)~107505, {\tt hep-ph/9907080};\\
W. Goldberger and M. Wise, 
{\it \lq \lq Modulus Stabilization with Bulk Fields,''}
Phys. Rev. Lett. {\bf 83}~(1999)~4922, {\tt hep-ph/9907447}.
\bibitem{fermi} 
S. Chang et al.,
{\it \lq \lq Bulk Standard Model in the Randall-Sundrum Background,''}
Phys. Rev. {\bf D62}~(2000)~084025, {\tt hep-ph/9912498}.
\bibitem{mass} 
T. Gherghetta and A. Pomarol,
{\it \lq \lq Bulk Fields and Supersymmetry in a Slice of AdS,''}
Nucl. Phys. {\bf B586}~(2000)~141, {\tt hep-ph/0003129}.
\bibitem{dv}
G. Dvali and M. Shifman,
{\it \lq \lq Domain Walls in Strongly Coupled Theories,''}
Phys. Lett. {\bf B396}~(1997)~64; Erratum-ibid. 
{\bf B407}~(1997)~452, {\tt hep-th/9612128}.
\bibitem{gauge} 
H. Davoudiasl, J. L. Hewett and T. G. Rizzo,
{\it \lq \lq Bulk Gauge Fields in the Randall-Sundrum Model,''}
Phys. Rev. Lett. {\bf 84}~(2000)~2080, {\tt hep-ph/9911262};\\
A. Pomarol, 
{\it \lq \lq Gauge Bosons in a Five-dimensional Theory 
with Localized Gravity,''}
Phys. Lett. {\bf B486}~(2000)~153, {\tt hep-ph/9911294};\\
S. Chang et al. in Ref.\cite{fermi}.
\bibitem{conformal}
M. Shaposhnikov and P. Tinyakov,
{\it \lq \lq Extra Dimensions as an Alternative to Higgs
Mechanism?,''}
{\tt hep-th/0102161};\\
M. Tachibana,
{\it \lq \lq Comment on Kaluza-Klein Spectrum of Gauge Fields in the 
Bigravity Model,''}
JHEP {\bf 0107}~(2001)~021, {\tt hep-th/0105180}. 
\bibitem{ghoroku} 
K. Ghoroku and A. Nakamura, 
{\it \lq \lq Massive Vector Trapping as a Gauge Boson on a Brane,''}
{\tt hep-th/0106145}.
\bibitem{kogan} 
I. I. Kogan, S. Mouslopoulos, A. Papazoglou and G. G. Ross, 
{\it \lq \lq Multi-Localization in Multi-Brane Worlds,''}
{\tt hep-ph/0107307}.
\bibitem{kehagias} 
A. Kehagias and K. Tamvakis, 
{\it \lq \lq Localized Gravitons, Gauge Bosons and Chiral Fermions
in Smooth Spaces Generated by a Bounce,''}
Phys. Lett. {\bf B504}~(2001)~38, {\tt hep-th/0010112}.
\bibitem{kawai}
H. Kawai and T. Kuroki,
{\it \lq \lq Strings as Flux Tube and Deconfinement on Branes
in Gauge Theories,''}
{\tt hep-th/0106103}.
\bibitem{thooft}
G. 't~Hooft, CERN preprint~(1974);\\
J. Kogut and L. Susskind,
{\it \lq \lq Vacuum Polarization and the Absence of Free Quarks
in Four Dimensions,''}
Phys Rev. {\bf D9}~(1974)~3501.
\end{thebibliography}
\end{document}